# Eddy Current Testing of Metal Cracks Using Spin Hall Magnetoresistance Sensor and Machine Learning

Yanjun Xu, Yumeng Yang, and Yihong Wu*

*Abstract* — Recently we have developed a spin Hall magnetoresistance (SMR) sensor which operates under AC bias and sense currents. Here we demonstrate both theoretically and experimentally that the SMR sensor is uniquely suited for eddy current testing applications because both the coil and sensor utilize AC current as the excitation source. The use of SMR sensor effectively eliminates the necessity of any demodulation or lock-in technique for detecting the eddy current, which greatly simplifies the detection system. Furthermore, we show that the combination of principal component analysis and decision tree model is effective in classifying the metal cracks. The relatively clean signals obtained by the SMR sensor greatly facilitates the subsequent signal analysis and ensures high accuracy in the classification of different types of crack features.

## I. INTRODUCTION

Non-destructive testing (NDT) is the process of inspecting materials or components for discontinuities or inhomogeneities without impairing function of the material[1], [2]. It has been widely used in manufacturing, fabrication and inspections in order to ensure product integrity and reliability. The most commonly used non-destructive techniques include electromagnetic, ultrasonic, X-ray, and liquid penetrant testing, among which the eddy current based NDT, or simply eddy current testing (ECT), has proven to be one of the most effective electromagnetic NDT techniques for the inspection of conductive materials such as copper, aluminium or steel [3]. In ECT, an excitation coil is excited by an alternating current (AC) to generate a time-varying magnetic field in the conductive test piece, which in turn induces an eddy current that resists the flux change in the coil. The presence of any defect or crack in the test piece would affect both the amplitude and distribution of the eddy current. The change in the eddy current can then be detected as a signal to pinpoint or map out the defects[3], [4].

There are several different types of sensors that can be used to detect the change of the eddy current induced by the cracks in test piece, including coil probe [5], superconducting quantum interference device (SQUID) [6]-[8], Hall-effect [9] and magnetoresistive (MR) sensors [10]-[15]. The coil probe generally has two types, namely, the single-coil and dual-coil probes. The former uses the same coil to generate the eddy current and at the same time to detect the impedance changes induced by the eddy current (Fig. 1(a)) [16]. In order to mitigate the effect of lift-off height, temperature change and material variation, typically differential probes are employed in practical designs. The primary drawback of single-coil probe is that the coil design can hardly be optimized for both the excitation and detection functions. The dual-coil probe, on the other hand, employs a primary coil to excite the eddy current and a separate secondary coil (or pick-up coil) to sense the field generated by the eddy current (Fig.1(b)). Compared to the single-coil probe, the dual-coil design allows to optimize the excitation coil for eddy current generation with maximum efficiency and the pick-up coil for detecting the secondary field with maximum sensitivity [17]. In addition, a differential coil design has been explored to improve the detection sensitivity, eliminate primary field and background noise [18], [19]. Despite the flexibility in coil minimization, the inductive detection technique carries the inherent drawback of decreasing sensitivity at low-frequency and poor spatial resolution as manifested in its signal-to-noise ratio (SNR), *i.e.*, $SNR \propto r^{5/2} f_c$, where $r$ is the coil radius and $f_c$ the operating frequency [20]. Although SQUID and Hall effect sensors can provide some partial solutions to the challenges faced by inductive sensors, both types of sensors have their own disadvantages, *e.g.*, SQUID can only operate at cryogenic temperature, and Hall-effect sensor suffers from limited sensitivity [13], [20].

The MR sensors, including anisotropic magnetoresistance (AMR), giant magnetoresistance (GMR) and tunnel magnetoresistance (TMR) sensors [21]-[24], are attractive for ECT applications due to their high sensitivity at low magnetic field over a broad frequency range and the ability to directly measure the magnetic field instead of its time derivative (Fig.1(c)). This makes it possible to realize ECT with high spatial resolution and large detection depth [10-15], [20]. Due to the limited dynamic range, the MR sensors have to be arranged in such a way that the influence of the primary field be minimized. This can be done by either placing the sensor with its sensing axis perpendicular to the primary field direction or using a compensation coil to cancel out the primary field. Regardless of the design, in most cases, the MR sensor is driven by a direct current (DC) and additional demodulation or lock-in technique is required to decouple the eddy current related signal from the overall signals, which is quite challenging.

This work was supported by Singapore Ministry of Education, under its AcRF Tier 2 Grant (Grant no. MOE2017-T2-2-011 and MOE2018-T2-1-076).

The authors are with the Department of Electrical and Computer Engineering, National University of Singapore, 4 Engineering Drive 3, Singapore 117583 (* Author to whom correspondence should be addressed: elewuyh@nus.edu.sg)

Accurate reconstruction of the crack features still remains elusive in these ECT systems because of the difficulties in calibration and the lack of system-level signal processing algorithms [25], [26]. Recently, we have developed a spin Hall magnetoresistance (SMR) sensor using the spin orbit torque (SOT) induced effective field ($H_{FL}$) as the built-in sensor linearization mechanism [27], [28]. The use of SOT biasing greatly simplifies the sensor design as it eliminates the needs of delicate transverse bias used in conventional AMR, GMR and TMR sensors. Another key difference between SMR and other types of MR sensors is that the SMR sensor is driven by an AC current, but the output is a DC voltage. The combination of SOT biasing and AC driving current makes it possible to realize an SMR sensor with extremely simple structure, high sensitivity, nearly zero dc offset, negligible hysteresis, and a detectivity around 1 nT/$\sqrt{Hz}$ at 1 Hz [29], [30]. The AC driving capability makes the SMR sensor uniquely suited for ECT applications as it does not require the use of lock-in technique to detect the signal (the detection principle will be presented in the next section). Specifically, we fabricate SMR sensors based on NiFe(2.5)/Au$_{19}$Pt$_{81}$(3.2) bilayers (the numbers inside the parenthesis are thicknesses in nm) and investigate the performance of the SMR sensor in detecting cracks with different dimensions in an aluminum (Al) plate. We choose NiFe(2.5)/Au$_{19}$Pt$_{81}$(3.2) bilayer because it exhibits a much lower power consumption and more than 80% enhancement of spin-orbit torque efficiency as compared to the NiFe/Pt bilayer sensor [30]. We demonstrate both theoretically and experimentally that the SMR sensor is promising for ECT without the necessity of any demodulator or lock-in amplifier. With the assistance of principal component analysis and decision tree model, cracks of different features are successfully detected using the SMR-based ECT probe.

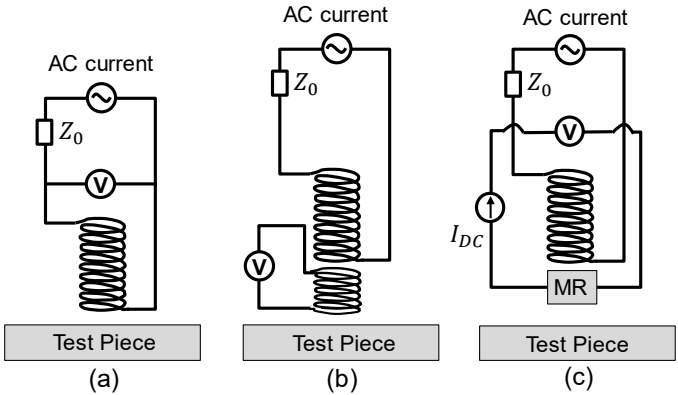

Fig.1. Schematic of ECT using: (a) single-coil, (b) dual-coil, and (c) MR sensors.

## II. THEORY OF EDDY CURRENT FIELD DETECTION USING SMR SENSOR

In this section, we explain how the SMR sensor can be used to detect the eddy current field. As shown schematically in Fig. 2(a), the SMR sensor consists of four NiFe(2.5)/Au$_{19}$Pt$_{81}$(3.2) bilayer elements with the same dimension which are arranged in the form of a Wheatstone bridge. The elements have an elliptical shape with the easy axis in the long-axis or x-direction and the hard axis in the short-axis or y-direction. When an AC current, $I = I_o \sin(\omega_s t + \theta_0)$, is applied to two terminals of the bridge, the output voltage across the other two terminals is given by [28]-[30]: (detailed derivation is give in Supplementary Note 1):

$$V_{out} = \frac{1}{2} I_o \Delta R_0 \sin(\omega_s t + \theta_0) + \frac{1}{2} \frac{\alpha I_0^2 \Delta R H_y \cos 2(\omega_s t + \theta_0)}{(H_d + H_k)^2} - \frac{1}{2} \frac{\alpha I_0^2 \Delta R H_y}{(H_d + H_k)^2} \quad (1)$$

where $I_o$ is the amplitude of the AC current applied to the SMR sensor, $\Delta R_o$ is the offset resistance between the neighbouring sensing elements, $\omega_s$ and $\theta_0$ are the angular frequency and phase of the sensor driving current, $\Delta R$ is the resistance change due to both the SMR and AMR when the magnetization changes from parallel to perpendicular to the current direction, $H_d$ is the in-plane demagnetizing field, $H_k$ is the uniaxial anisotropy field, and $H_y$ is the external magnetic field or field to be detected in y-direction. In the case of NiFe/Au$_{19}$Pt$_{81}$ bilayer, the applied current $I$ mainly flows through the Au$_{19}$Pt$_{81}$ layer due to its much smaller resistivity, which generates both a field-like effective field $H_{FL}$ and an Oersted field $H_{Oe}$ in the NiFe layer, with $H_{FL}$ and $H_{Oe}$ in the same direction and $H_{FL} \gg H_{Oe}$. In (1), $\alpha = (H_{FL} + H_{Oe})/I$, indicates the efficiency of SOT generation in the NiFe/Au$_{19}$Pt$_{81}$ bilayers. When the sensor is used to detect a static field, the last term of (1) gives a DC output signal proportional to the external field. The first two terms don't contribute to the DC output signal as the time-average of both terms are zero. The sensor is linear as long as the external field does not exceed the dynamic range, which is mostly determined by the shape anisotropy of the sensor. Fig. 2(b) shows the typical static field response of a SMR sensor based on NiFe(2.5)/Au$_{19}$Pt$_{81}$(3.2) bilayers. The sensor was fabricated using combined techniques of sputtering/evaporation and lift-off. The NiFe layer was deposited by sputtering and the Au$_{19}$Pt$_{81}$ layer was deposited by evaporation. Both layers were deposited in a multi-chamber system at a base pressure below $3 \times 10^{-8}$ Torr without breaking the vacuum. An in-plane field of ~500 Oe was applied along the long axis of sensing elements during the deposition to induce a uniaxial anisotropy for the NiFe layer. The root mean square (rms) amplitude and frequency ($f_s$) of the applied AC current density are $9.4 \times 10^5$ A/cm$^2$ and 5000 Hz, respectively. The response curve is obtained by sweeping the field in y-direction from -2 Oe to +2 Oe and then back to -2 Oe. It can be seen that the forward and backward sweeping response curves nearly overlap with each other, indicating a negligible hysteresis in the full field range. In addition, the DC offset is also nearly zero. Within the linear range of ±0.84 Oe, the sensor exhibits a sensitivity of around 1.10 mV/V/Oe. When the SMR sensor is used to detect the eddy current in ECT, the external field is composed of both the primary field from the excitation coil and the secondary field from the eddy current. Hence the external field to be detected in y-direction can be expressed as: $H_y = H_c \sin \omega_c t + \beta H_e \sin(\omega_c t + \frac{\pi}{2} + \theta_L)$, where $H_c$ ($H_e$) is the amplitude of the primary (eddy current) field, $\omega_c$ is the angular frequency of the AC current applied to the excitation coil, $\beta$ is a parameter quantifying the influence of cracks on the eddy current ($\beta = 1$ for crack-free case and $0 < \beta < 1$ for the case with crack), and $\theta_L$ is the phase lag between surface and subsurface eddy current (it is a function of crack depth from the surface). It is worth noting that the $\frac{\pi}{2}$ term is the phase difference between the primary field and the eddy current field on the test piece surface. Without losing generality, we assume that there is a phase difference of $\theta_0$ between the driving current for the excitation coil and the sensor, i.e., $I = I_0 \sin(\omega_s t + \theta_0)$ for the sensor. By substitute $H_y$ into (1), we obtain

$$V_{out} = \frac{I_0 \sin(\omega_s t + \theta_0)}{2} \Delta R_0 + \frac{1}{2} \alpha I_0^2 \Delta R \cos 2(\omega_s + \theta_0)$$
$$\frac{H_c \sin \omega_c t + \beta H_e \sin\left(\omega_c t + \frac{\pi}{2} + \theta_L\right)}{(H_k + H_d)^2}$$
$$-\frac{1}{2}\alpha I_{AC}^2 \Delta R \frac{H_c \sin \omega_c t + \beta H_e \sin(\omega_c t + \frac{\pi}{2} + \theta_L)}{(H_k + H_d)^2}. \quad (2)$$

The 1st and 3rd terms are AC components, whereas the 2nd term produces a DC output when $\omega_c = 2\omega_s$, i.e.,

$$V_{DC} = -\frac{1}{4}\alpha I_{AC}^2 \Delta R \frac{1}{(H_k + H_d)^2}\left\{H_c \sin 2\theta_0 + \beta H_e \left[\sin 2(\theta_0 - \frac{\pi}{4} - \frac{\theta_L}{2})\right]\right\}. \quad (3)$$

This DC output voltage contains two terms. The first term is due to the primary field generated by the excitation coil and the second term is from the eddy current induced secondary field. The second term can be used to detect the cracks as it contains the parameter $\beta$. The phase difference between these two signals is $\frac{\pi}{2} + \theta_L$. When the crack is on the surface, $\theta_L \approx 0$, leading to a $\frac{\pi}{2}$ phase difference between the primary field and secondary field for surface cracks. (See Supplementary Note 2 for detail explanation)

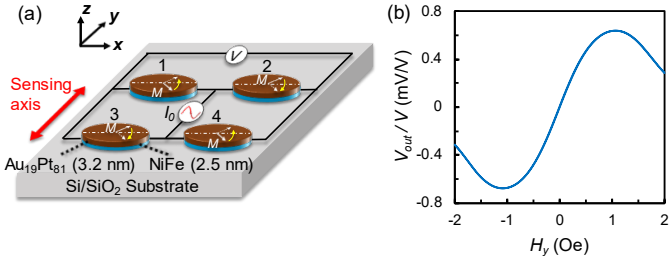

Fig. 2. (a) Schematic of a Wheatstone bridge SMR sensor comprising of four ellipsoidal NiFe(2.5)/Au$_{19}$Pt$_{81}$(3.2) bilayer sensing elements with the arrows indicating the magnetization direction driven by the AC current. The sensing axis is indicated by red arrows. (b) Response curve of the sensor obtained by sweeping the field in y-direction from -2 Oe to +2 Oe and then back to -2 Oe. The rms amplitude and frequency of the applied AC bias current density are $9.4\times10^5$ A/cm$^2$ and 5000 Hz, respectively.

## III. EXPERIMENTAL DETAILS

The ECT setup consists of a customized ECT probe, AC current sources for the excitation coil and SMR sensor, Al plate with pre-formed cracks, a DC nanovoltmeter and a personal computer for data acquisition and processing. As shown schematically in Fig. 3(a), during the crack detection experiment, the ECT probe is placed at a fixed height, whereas the Al test piece is moved along x-direction with a constant speed of 5.4 mm/s driven by a linear motor. The excitation coil generates an eddy current in the Al plate and the SMR sensor detects the field generated by both the excitation coil and the eddy current. As shown in Fig. 3(b), the customized ECT probe consists of an excitation coil with an inner diameter of 12 mm, and height of 15 mm. The number of turns is 100 and the diameter of the wire is 0.2 mm. The SMR sensor is placed on a PCB board, which is inserted inside the coil. The sensing axis is along the axial direction of the coil. The sensors are placed as close as possible to the bottom edge of the coil so as to improve the signal-to-noise ratio. Five different types of cracks characterized by its depth ($z$), height ($h$) and width ($w$) are pre-formed on the Al plate (see Fig. 3(a) for the definition of crack features). Detailed dimensions of the cracks are summarized in Table 1. Among them, cracks 2-4 are surface cracks and crack 1 is a subsurface crack. For each crack, 45 measurements are carried out by repeating the same scanning process over the region with crack using the same scanning speed. Each measurement generates a voltage-time series of data with 141 data points (the interval for each step is 0.071 s).

TABLE 1. Summary of five types of cracks with different dimensions and locations. The thickness of all the Al plates is 5 mm and the crack length along the crack direction is fixed at 10 cm. Crack photos are shown in Supplementary Fig. (S3).

|  | Crack width $w$ (mm) | Crack Height $h$ (mm) | Crack depth $z$ (mm) | No. of measurements |
|---|---|---|---|---|
| Crack-free | 0 | 0 | 0 | 45 |
| Crack 1 (subsurface) | 5 | 2 | 3 | 45 |
| Crack 2 (surface) | 1 | 2 | 0 | 45 |
| Crack 3 (surface) | 2 | 2 | 0 | 45 |
| Crack 4 (surface) | 5 | 2 | 0 | 45 |

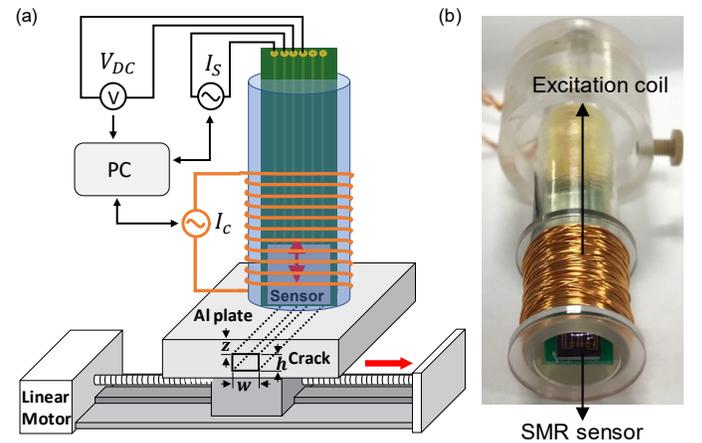

Fig. 3 (a) Schematic of the ECT experiment setup which consists of a fixed ECT probe and a moving test piece driven by a linear motor. (b)

## IV. RESULTS AND DISCUSSION

### A. Crack detection using the SMR sensor

To test the performance of the SMR-based ECT probe, we conducted the aforementioned measurements on crack 3 (surface crack with a height of 2 mm and width of 2 mm). In the measurement, the excitation coil is driven by an AC current with an amplitude of 40 mA and frequency ($f_c$) of 1000 Hz, whereas the SMR sensor is driven by an AC current with an amplitude density of $9.4 \times 10^5$ A/cm$^2$ and frequency ($f_s$) of 500 Hz. The amplitude and frequency for the driving current for both the coil and sensor are fixed, unless otherwise specified. The SMR sensor output is measured by a nanovoltmeter. Fig. 4(a) summarized the results obtained in three separate measurements for the same crack with different phase difference between the driving current for the excitation coil and the sensor ($\theta_0 = 0$, 45º, 90º, respectively). The curves are shifted vertically for clarity. The crack induced signal change can be seen clearly in the middle of the data series ($V_{Crack}$). A positive peak appears in the crack region when $\theta_0 = 0$, which nearly flattens out at $\theta_0 = 45$º, and then reveres its polarity at $\theta_0 = 90$º. The baseline ($V_{Base}$) corresponds to the primary field which varies with $\theta_0$ as well. These results are in good agreement with (3). When $\theta_0 = 0$, the first term of (3) which corresponds to the output of the primary field becomes zero, while the second term which is related to the eddy current becomes positive maximum; when $\theta_0 = 45$º, the primary field output is maximum, but the eddy current term becomes nearly zero; when $\theta_0 = 90$º, the primary field contribution becomes minimum again, but the eddy current related term becomes negative maximum. In order to have a more quantitative understanding of the phase dependence of the output DC signal, we measure $V_{Base}$ and $V_{Crack}$ as a function of $\theta_0$ in the range of 0 to $\pi$ and the results are shown in Fig. 4(b). As expected from (3), both $V_{Base}$ and $V_{Crack}$ follow sinusoidal relation with respect to $\theta_0$, while the phase difference between $V_{Base}$ and $V_{Crack}$ is $\frac{\pi}{2}$. All these results are correlated well with the theoretical model described in Section II, and they demonstrate clearly the advantages of using SMR sensor for ECT applications. The signal detection without any demodulation circuits or lock-in amplifier greatly simplifies the design of the ECT system.

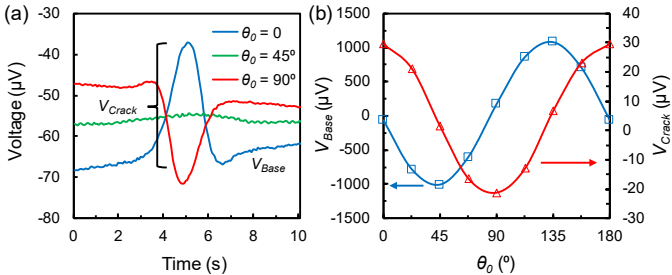

Fig. 4. (a) Output of the SMR sensor when the ECT probe is scanned over crack 3. The three curves are obtained by setting $\theta_0 = 0$, 45º, 90º, respectively. The curves are shifted vertically for clarity. $V_{Base}$ and $V_{Crack}$ corresponds to the primary and eddy current field, respectively. (b) $\theta_0$ dependence of $V_{Base}$ and $V_{Crack}$ in the range of 0 to $\pi$.

### B. ECT measurements on different types of cracks

Next, we use the same ECT setup and measurement condition to detect different types of cracks listed in Table I by fixing $\theta_0$ at 0. The measurement of each type of crack is repeated for 45 times. Fig. 5(a) shows a typical measurement result for the crack-free case. The slightly fluctuating signal corresponds to $V_{Base}$. As expected, there is no observable features in the repeated measurements for the crack-free test piece. Fig. 5(b) to (e) show the five consecutive measurement results of cracks 2 - 4. The curves have been shifted vertically for clarity. In a sharp contrast to the crack-free case, peaks corresponding to different types of cracks are clearly seen and the results are reproducible in different runs of measurements. It is not surprising that the signals from surface cracks (cracks 2-4) are obviously stronger than that of the subsurface crack (crack 1). The smaller signal of subsurface crack can be accounted for by the attenuation effect as the crack is father away from both the excitation coil and the sensor. In addition, the phase lag of the eddy current field may also play a role as $\theta_0$ is set at 0 for the surface cracks. Fig. 5(f) compares the output signals from different cracks including the crack-free case (note: all the data have been centralized and z-normalized, which will be explained shortly). Although both the amplitude and width of the broad peak vary among different types of cracks, they cannot be directly used to discriminate different crack features because both the amplitude and shape of the crack related signal can be easily affected by the measurement conditions and there is a lack of suitable signal processing algorithm to remove such fluctuations [25]. Therefore, after obtaining the time series of measurement results for all types of cracks, we perform principal component analysis (PCA) to extract the important principal components and then use the decision tree model to differentiate different types of crack features.

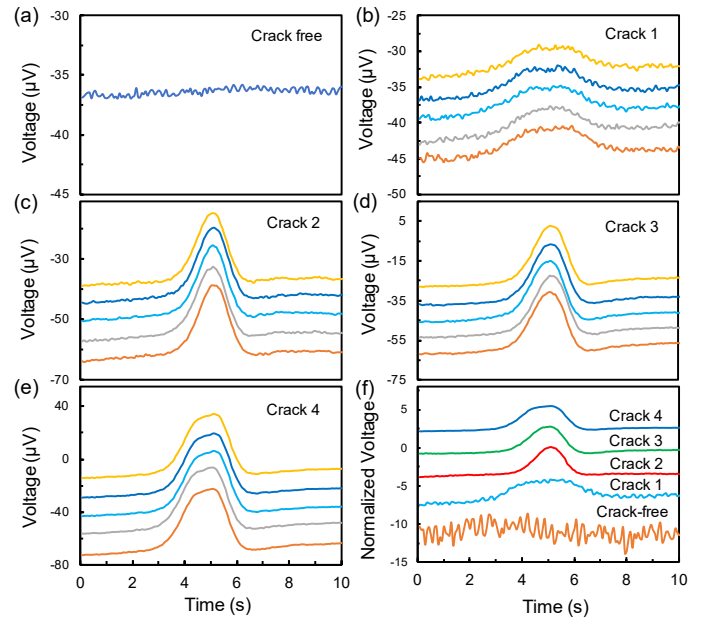

Fig.5. (a) Typical SMR sensor output signal of the crack-free test piece. (b) - (e) Output signals of five repeated ECT measurements for cracks 1 – 4, respectively. All the curves have been shifted vertically for clarity. (f) Centralized and z-normalized output data for different types of cracks including the crack-free case. Again the curves are shifted along the vertical axis.

### C. Crack classification using principal component analysis and decision tree model

The PCA is a statistical signal processing technique that uses an orthogonal transformation to convert a set of observations of possibly correlated variables into a set of values of linearly uncorrelated variables called principal components (PC) [31],[32]. It is widely used to extract dominant features from a set of multivariate date including in NDE [33]-[36]. The typical procedure of PCA involves the following: 1) preparation of the data to make it suitable for PCA by removing the mean and normalizing it by the standard deviation, 2) computing the eigenvectors and eigenvalues of the covariance matrix of the pre-processed data, 3) sorting the eigenvectors in descending order of the corresponding eigenvalues, 4) constructing the projection matrix by combining the sorted eigenvectors, and 5) transforming the original dataset using the projection matrix to obtain a new dataset, called scoring matrix; the entries of the scoring matrix can be considered as the projection of the original dataset in the new feature space spanned by the principal components. The eigenvectors determine the directions of the principal component axes, and the corresponding eigenvalues indicate the variance of the original data in that direction. The first PC has the largest eigenvalue, hence largest variance, followed by other PCs in descending order of eigenvalues. As the original dataset can be reproduced using the first few PCs, it effectively reduces the dimensionality of the original data by retaining only those which give the largest variance.

In the present case, the datasets are the time-domain ECT measurement data of different cracks. As can be seen from the measured signals, the dataset obtained from different cracks have a high degree of similarity, which makes it difficult to distinguish the cracks from each other. By performing the PCA, we can effectively reduce the similarity and extract the components with largest variance representing the key features of each test piece. Depending on the number of PCs that are required to construct the original data, one may also leverage on other data analysis techniques such as neural network and decision tree analysis to classify the test features. Fig. 6(a) is the block diagram of the PCA and decision tree analysis adopted in this work. The first step is to prepare the data and make them suitable for PCA analysis. To this end, we scale the time series dataset to zero mean by subtracting the mean value and then normalized it by the respective standard deviation (z-normalization). Fig. 5(f) shows the measured data for different types of cracks after the mean subtraction and z-normalization. We can see that all the signals are now at comparable scale and suitable for further processing. The pre-processed data then are combined into a matrix

$$X = \begin{bmatrix} x_{t1}^1 & \cdots & x_{tn}^1 \\ \vdots & \ddots & \vdots \\ x_{t1}^m & \cdots & x_{tn}^m \end{bmatrix} \quad (4)$$

where $t1$ to $tn$ refer to the time steps in a particular measurement and $m$ is the number of measurements. In the present case, $m$ is 225, comprising of 45 measurements for each type of cracks and $n$ is 141. The next step is to perform PCA on $X$, in which eigenvalues and eigenvectors of the covariance matrix of $X$ are calculated using the Matlab. After arranging the eigenvectors based on the descending order of the corresponding eigenvalues, the projection matrix $W$ with dimension of 141 × 141 is constructed by combining the sorted column eigenvectors (with $i$th column corresponding to $i$th PC). Fig. 6(b) shows the eigenvectors of the first 3 PCs. As the difference in eigenvalues corresponding to the first few PCs is not very large, we need more PCs to distinguish different types of cracks. Fig. 6(c) shows the eigenvalues of the first 15 principal components. Due to the nature of the problem at hand, the cumulative variance increases relatively slowly with the number of PCs, and at the 15th PC, the cumulative variance is about 80%. As a trial, we use the first 15 PCs to reconstruct the original dataset. To this end, we calculate the scoring matrix $P = X \cdot W$ and the centered scoring matrix

$$P_c = P - \overline{P} = \begin{bmatrix} PC1^1 & \cdots & PCn^1 \\ \vdots & \ddots & \vdots \\ PC1^m & \cdots & PCn^m \end{bmatrix} \quad (5)$$

where $\overline{P}$ is the column average of $P$, and $PCk^l$ is the inner product of $k$th principal component and the $l$th measurement data. After obtaining the scoring matrix, the first 15 entries of each row of $P_c$, are used to classify different types of cracks based on decision tree analysis.

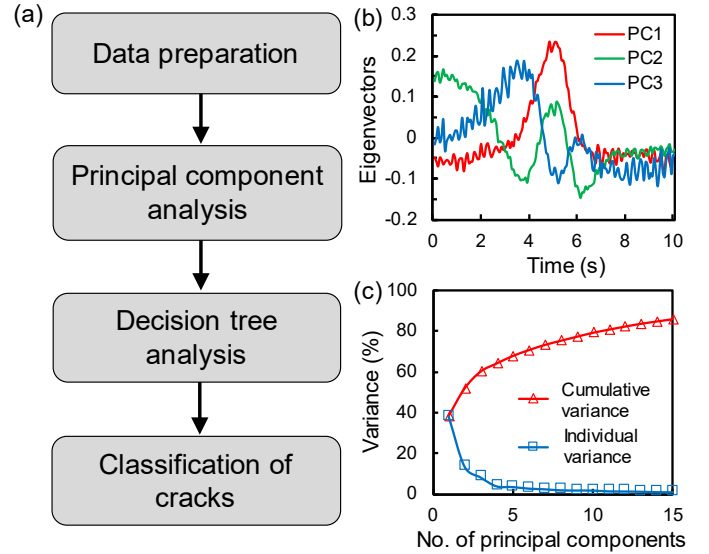

Fig.6. (a) Block diagram of crack classification using the PCA and decision tree analysis. (b) Eigenvectors of the first 3 PCs. (c) Cumulative and individual variance for the first 15 principal components.

The centered scoring matrix $P_c$ has a dimension of 225 × 141. Each row corresponds to the projection of the original dataset in the 141 principal component axes, among which the values of the first 15 entries, *i.e.*, $PCk^l$, $k = 1$ to 15, are representative of the original datasets. Therefore, a decision tree model can be built based on the 15 $PCk^l$ ($k = 1$ to 15 and $l = 1$ to 225) values to distinguish the type of crack in the $l$th measurement. Fig. 7(a) is a plot of the 2$^{nd}$ principal component projection $PC2^l (l = 1$ to 225) against that of the 1$^{st}$ one $PC1^l (l = 1$ to 225), which is able to distinguish crack-free (diamond), subsurface crack 1 (cross) and surface crack 4 (circle), as indicated by the dotted line. However, surface cracks 2 and 3 can hardly be differentiated because of their similar crack features. The only difference is that crack 2 has a width of 1 mm and crack 3 has a width of 2 mm. Interestingly the two types of cracks can be clearly distinguished by using the 10$^{th}$ PC projection value of 0.15, as shown by the dotted line in Fig. 7(b). More specifically, all the different types of cracks can be classified with 100% accuracy using the decision tree model as shown in Fig. 7(c). The classification flowchart is as follows. First, we set the condition of $PC1^l < -1.01$; this effectively separates the crack-free test pieces from the rest with cracks. Second, we can further distinguish subsurface crack from the surface cracks by using the criterion $PC2^l < -2$. Lastly, the three surface cracks can be differentiated by first requiring $-2 < PC2^l < 0.67$ to separate crack 4 with largest width from the two narrower cracks, and then using $PC10^l$ to classify crack 2 and 3. The combination of PCA and decision tree model is able to classify the cracks with 100% accuracy.

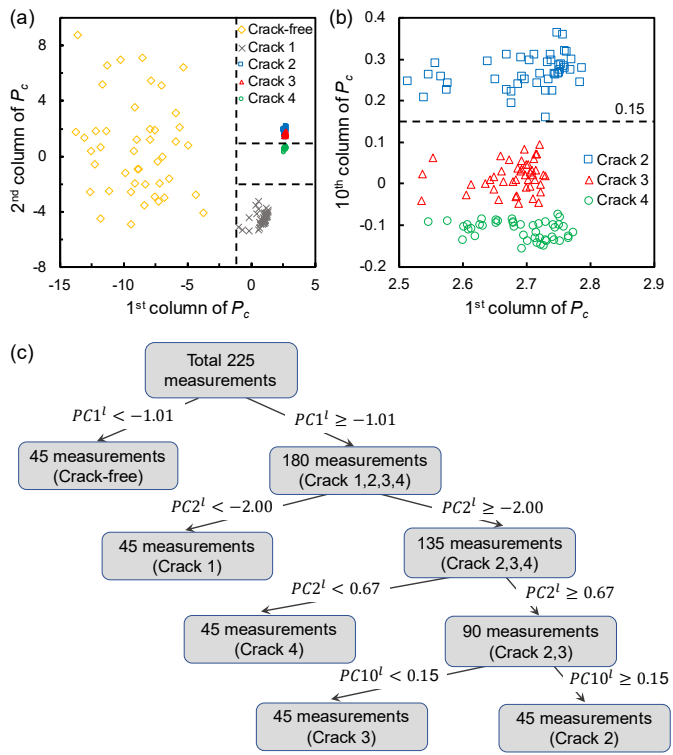

Fig.7. (a) Plot of the 2$^{nd}$ column against the 1$^{st}$ column of the centred scoring matrix. The dotted lines help distinguish crack-free (diamond), subsurface crack 1 (cross) and surface crack 4 (circle) of the test pieces. (b) Plot of the 10$^{th}$ column against the 1$^{st}$ column of the centred scoring matrix. The dotted line separates surface crack 2 (square) from crack 3 (triangle). (c) Illustration of the decision tree model to classify all the different types of cracks ($l = 1$ to 225).

## V. CONCLUSIONS

In summary, we have demonstrated, both theoretically and experimentally, that the SMR sensor is promising for ECT without the necessity of any demodulator or lock-in amplifier. This is possible because both the excitation coil and the SMR sensor are driven by AC current; the built-in rectification effect gives a DC output which facilitates data acquisition and processing. The nearly zero hysteresis and DC offset help to reduce the fluctuations in the measured data. Furthermore, we have shown that the combination of PCA and decision tree model is effective in distinguishing different types of cracks. The relatively clean data obtained by the SMR sensor greatly facilitates the subsequent PCA analysis and ensures high accuracy in classification of different crack features.

REFERENCES

[1] B. Raj, T. Jayakumar, and M. Thavasimuthu, *Practical non-destructive testing*: Woodhead Publishing, 2002.
[2] J. H. Charles, *Handbook of Nondestructive Evaluation, Second Edition*, 2$^{nd}$ ed., New York: McGraw-Hill Education, 2013.
[3] L. Janousek, K. Capova, N. Yusa, and K. Miya, "Multiprobe inspection for enhancing sizing ability in eddy current nondestructive testing," *IEEE Transactions on Magnetics,* vol. 44, no. 6, pp. 1618-1621, May 2008.
[4] H. Hashizume, Y. Yamada, K. Miya, S. Toda, K. Morimoto, Y. Araki, K. Satake, and N. Shimizu, "Numerical and experimental analysis of eddy current testing for a tube with cracks," *IEEE transactions on magnetics,* vol. 28, no. 2, pp. 1469-1472, Mar. 1992.
[5] S. Tumanski, "Induction coil sensors - A review," *Measurement Science and Technology,* vol. 18, no. 3, pp. R31, Jan. 2007.
[6] D. He, M. Daibo, and M. Yoshizawa, "Mobile HTS rf SQUID magnetometer," *IEEE transactions on applied superconductivity,* vol. 13, no. 2, pp. 200-202, Jul. 2003.
[7] Y. Tavrin, H.-J. Krause, W. Wolf, V. Glyantsev, J. Schubert, W. Zander, and H. Bousack, "Eddy current technique with high temperature SQUID for non-destructive evaluation of non-magnetic metallic structures," *Cryogenics,* vol. 36, no. 2, pp. 83-86, 1996.
[8] D. He, and M. Yoshizawa, "Dual-frequency eddy-current NDE based on high-Tc rf SQUID," *Physica C: Superconductivity,* vol. 383, no. 3, pp. 223-226, Dec. 2002.
[9] D.-G. Park, C. S. Angani, B. Rao, G. Vértesy, D.-H. Lee, and K.-H. Kim, "Detection of the subsurface cracks in a stainless steel plate using pulsed eddy current," *Journal of Nondestructive Evaluation,* vol. 32, no. 4, pp. 350-353, Jul. 2013.
[10] A. Bernieri, G. Betta, L. Ferrigno, and M. Laracca, "Improving performance of GMR sensors," *IEEE Sensors Journal,* vol. 13, no. 11, pp. 4513-4521, Jun. 2013.
[11] G. Betta, L. Ferrigno, and M. Laracca, "GMR-based ECT instrument for detection and characterization of crack on a planar specimen: A hand-held solution," *IEEE transactions on instrumentation and measurement,* vol. 61, no. 2, pp. 505-512, Aug. 2011.

[12] K. Allweins, M. Von Kreutzbruck, and G. Gierelt, "Defect detection in aluminum laser welds using an anisotropic magnetoresistive sensor array," *Journal of applied physics,* vol. 97, no. 10, pp. 10Q102, 2005.

[13] D. He, and M. Shiwa, "Deep Defect Detection Using Eddy Current Testing with AMR Sensor," *Progress In Electromagnetics Research Symposium Proceedings,* Stockholm, Sweden, 2013, pp. 493-495.

[14] D. Rifai, A. N. Abdalla, K. Ali, and R. Razali, "Giant magnetoresistance sensors: A review on structures and non-destructive eddy current testing applications," *Sensors,* vol. 16, no. 3, pp. 298, Feb. 2016.

[15] K. Tsukada, M. Hayashi, Y. Nakamura, K. Sakai, and T. Kiwa, "Small eddy current testing sensor probe using a tunneling magnetoresistance sensor to detect cracks in steel structures," *IEEE Transactions on Magnetics,* vol. 54, no. 11, pp. 1-5, Jun. 2018.

[16] J. García-Martín, J. Gómez-Gil, and E. Vázquez-Sánchez, "Non-destructive techniques based on eddy current testing," *Sensors,* vol. 11, no. 3, pp. 2525-2565, Feb. 2011.

[17] A. L. Ribeiro, F. Alegria, O. Postolache, and H. G. Ramos, "Eddy current inspection of a duralumin plate," in *2009 IEEE Instrumentation and Measurement Technology Conference*, Singapore, 2009, pp. 1367-1371.

[18] L. S. Rosado, T. G. Santos, P. M. Ramos, P. Vilaça, and M. Piedade, "A differential planar eddy currents probe: Fundamentals, modeling and experimental evaluation," *NDT & E International,* vol. 51, pp. 85-93, Oct. 2012.

[19] G. Piao, J. Guo, T. Hu, Y. Deng, and H. Leung, "A novel pulsed eddy current method for high-speed pipeline inline inspection," *Sensors and Actuators A: Physical,* vol. 295, pp. 244-258, Aug. 2019.

[20] A. Jander, C. Smith, and R. Schneider, "Magnetoresistive sensors for nondestructive evaluation." in *Proc. SPIE 5770, Advanced Sensor Technologies for Nondestructive Evaluation and Structural Health Monitoring*, San Diego, CA, United States, 2005, pp. 1-13.

[21] K. Kuijk, W. v. Gestel, and F. Gorter, "The barber pole, a linear magnetoresistive head," *IEEE Transactions on Magnetics,* vol. 11, no. 5, pp. 1215-1217, Sep. 1975.

[22] S. Tumanski, *Thin film magnetoresistive sensors*, CRC Press, 2001.

[23] Y. Wu, *Nano Spintronics for Data Storage in Encyclopedia of Nanoscience and Nanotechnology*, Stevenson Ranch: American Scientific Publishers, 2003.

[24] P. Freitas, R. Ferreira, S. Cardoso, and F. Cardoso, "Magnetoresistive sensors," *Journal of Physics: Condensed Matter,* vol. 19, no. 16, pp. 165221, Apr. 2007.

[25] J. R. Bowler, and D. Harrison, "Measurement and calculation of transient eddy-currents in layered structures," in *Review of Progress in Quantitative Nondestructive Evaluation*, Springer, 1992, pp. 241-248.

[26] J. Cox, and D. Brown, "Eddy Current Instrumentation," *Zetec, Inc., Snoqualmie (WA), USA*, 2003.

[27] Y. Yang, Y. Xu, H. Xie, B. Xu, and Y. Wu, "Semitransparent anisotropic and spin Hall magnetoresistance sensor enabled by spin-orbit torque biasing," *Applied Physics Letters,* vol. 111, no. 3, pp. 032402, Jul. 2017.

[28] Y. Xu, Y. Yang, Z. Luo, B. Xu, and Y. Wu, "Macro-spin modeling and experimental study of spin-orbit torque biased magnetic sensors," *Journal of Applied Physics,* vol. 122, no. 19, pp. 193904, Nov. 2017.

[29] Y. Xu, Y. Yang, M. Zhang, Z. Luo, and Y. Wu, "Ultrathin All-in-One Spin Hall Magnetic Sensor with Built-In AC Excitation Enabled by Spin Current," *Advanced Materials Technologies,* vol. 3, no. 8, pp. 1800073, Jul. 2018.

[30] Y. Xu, Y. Yang, H. Xie, and Y. Wu, "Spin Hall magnetoresistance sensor using $Au_xPt_{1-x}$ as the spin-orbit torque biasing layer," *Applied Physics Letters,* vol. 115, no. 18, pp. 182406, Oct. 2019.

[31] V. Kshirsagar, M. Baviskar, and M. Gaikwad, "Face recognition using Eigenfaces." in *2011 3rd International Conference on Computer Research and Development*, Shanghai, 2011, pp. 302-306.

[32] I. Jolliffe, *Principal component analysis*, Springer, 2011.

[33] A. Sophian, G. Y. Tian, D. Taylor, and J. Rudlin, "A feature extraction technique based on principal component analysis for pulsed eddy current NDT," *NDT & e International,* vol. 36, no. 1, pp. 37-41, Jan. 2003.

[34] G. Tian, A. Sophian, D. Taylor, and J. Rudlin, "Wavelet-based PCA defect classification and quantification for pulsed eddy current NDT," *IEE Proceedings-Science, Measurement and Technology,* vol. 152, no. 4, pp. 141-148, Jul. 2005.

[35] P. Horan, P. Underhill, and T. Krause, "Pulsed eddy current detection of cracks in F/A-18 inner wing spar without wing skin removal using Modified Principal Component Analysis," *NDT & E International,* vol. 55, pp. 21-27, Apr. 2013.

[36] P. Zhu, Y. Cheng, P. Banerjee, A. Tamburrino, and Y. Deng, "A novel machine learning model for eddy current testing with uncertainty," *NDT & E International,* vol. 101, pp. 104-112, Jan. 2019.